\documentclass{PoS}

\usepackage{xspace}
\usepackage{amsmath}
\usepackage{lineno}

\newcommand{\jpsi}{\ensuremath{\text{J}/\psi}\xspace}
\newcommand{\Ups}{\ensuremath{\Upsilon}\xspace}

\newcommand{\pp}{\ensuremath{\rm pp}\xspace}

\newcommand{\PbPb}{\ensuremath{\text{Pb--Pb}}\xspace}
\newcommand{\XeXe}{\ensuremath{\text{Xe--Xe}}\xspace}
\newcommand{\s}{\ensuremath{\sqrt{s}}\xspace}
\newcommand{\snn}{\ensuremath{\sqrt{s_{\rm NN}}}\xspace}

\newcommand{\pt}{\ensuremath{p_{\text{T}}}\xspace}
\newcommand{\pT}{\ensuremath{p_{\text{T}}}\xspace}

\newcommand{\RAA}{\ensuremath{R_{\text{AA}}}\xspace}

\newcommand{\TAAmean}{\ensuremath{\langle T_{\text{AA}} \rangle}\xspace}

\newcommand{\Npart}{\ensuremath{N_{\text{part}}}\xspace}

\newcommand{\GeVc}{\ensuremath{{\rm GeV}/c}\xspace}

\newcommand{\der}{\ensuremath{\text{d}}\xspace}

\title{Quarkonium production in Pb--Pb and Xe--Xe collisions with ALICE at the LHC}

\ShortTitle{Quarkonium production in Pb--Pb and Xe--Xe collisions with ALICE at the LHC}

\author{\speaker{Markus K. K\"{o}hler}\thanks{on behalf of the ALICE Collaboration}\\
        Physikalisches Institut, Ruprecht-Karls-Universit\"{a}t Heidelberg, Germany\\
        E-mail: \email{markus.konrad.kohler@cern.ch}}

\abstract{Quarkonium production is an excellent probe to investigate the properties of the hot and dense medium, which can be created in heavy-ion collisions. The production and hadronisation of heavy quarks are well separated in the space-time evolution of a collision and provide a wealth of information of the underlying QCD dynamics from the dense to the eventually dilute system.

We report on the latest results from ALICE at the LHC on quarkonium production in heavy-ion collisions at mid- and forward rapidity. The nuclear modification factor as a function of centrality, transverse momentum, and rapidity is presented for charmonium and bottomonium states in $\PbPb$ collisions at $\snn = 5.02$~TeV. In addition, results on the nuclear modification factor of $\jpsi$ in $\XeXe$ collisions at $\snn=5.44 $~TeV are presented. Experimental results are compared with available phenomenological calculations.
}

\FullConference{International Conference on Hard and Electromagnetic Probes of High-Energy Nuclear Collisions\\
		30 September - 5 October 2018\\
		Aix-Les-Bains, Savoie, France}

\begin{document}


Quarkonium production is a fruitful probe for the investigation of the interaction between heavy quarks in the hot and energy-dense medium of heavy-ion collisions, as well as in the vacuum case in proton-proton ($\pp$) collisions, see e.g.~\cite{Brambilla2011,Andronic2016}. The production of the initial heavy-quark pair and the eventual formation of a hadronic bound state are well separated within the space-time evolution of a collision. This allows for a factorisation approach in $\pp$ collisions~\cite{Bodwin1995} and---due to the approximate conservation of the number of charm and beauty pairs throughout the evolution of a heavy-ion collision~\cite{Andronic2007}---for the investigation of the modification of quarkonium production in the medium produced in a heavy-ion collisions.

It was pointed out that the production of $\jpsi$ should be suppressed by the screening of charm quarks by the surrounding colour-dense medium~\cite{MatsuiSatz:1986}. This was confirmed by measurements at SPS and RHIC energies, whereas at LHC energies the suppression was observed to be significantly weaker, see e.g.~\cite{Arnaldi_2016} for a compilation of experimental results. The evolution of the suppression with increasing collision energy was predicted due to an additional production mechanism: the (re)generation of $\jpsi$ by the copiously produced charm quarks at LHC collision energies~\cite{BraunMunzinger:2000px,Thews:2000rj}. Models including (re)generation describe the majority of charmonium measurements from LHC to lower collision energies~\cite{Andronic2007,ZhaoRapp:2011,Ferreiro:2012rq} and establish (re)generation as the dominant production mechanism of $\jpsi$ at LHC energies. In addition, the study of bottomonium, for which the contribution from (re)generation could be small due to the larger mass of beauty quarks, can shed further light on the quarkonium production.

The study of \XeXe collisions provides a systematic cross check on quarkonium production: at a fixed number of nucleons participating in the collision, $\Npart$, the fireball volume of a \XeXe collision is expected to be comparable to the one of a \PbPb collision but with a different system geometry, see e.g.~\cite{ALICE_XeXe_pt,ALICE_XeXe_rapidity, ALICE_XeXe_flow}. In this way, the impact of the collision geometry on the quarkonium production and its dynamics can be studied.

%
%
ALICE~\cite{ALICE_detector2008,ALICE_detector2014} measures quarkonium production in the dielectron decay channel at mid-rapidity ($|y|<0.9$) and in the dimuon decay channel at forward rapidity ($2.5 < y < 4$). Events fulfilling the minimum-bias trigger condition are analysed for the dielectron analysis, where particles are identified as electrons via their specific energy loss in the Time Projecion Chamber. At forward rapidity, events are triggered and analysed in the muon spectrometer.

The nuclear modification factor $\RAA$ is used to quantify the influence of the medium on the particle production and its kinematics. It can be defined as 
\begin{equation}
  \RAA^{(X)} = \frac{\der^2 N^{\text{AA}}_X / \der y \der \pT}{\TAAmean \der^2\sigma^{\pp}_X / \der y \der \pT},
\end{equation}
where $\der^2 N^{\text{AA}}_X / \der y \der \pT$ is the efficiency corrected differential yield in nucleon-nucleon collisions for the particle species $X$, $\TAAmean$ is the centrality dependent nuclear overlap function and $\der^2\sigma^{\pp}_X / \der y \der \pT$ is the cross section measured in $\pp$ collisions at the same collision energy. A deviation from unity in the nuclear modification factor points to an influence of cold or hot nuclear matter on the particle production.

\begin{figure}[!t]
  \centering
  \includegraphics[width = 7.3 cm]{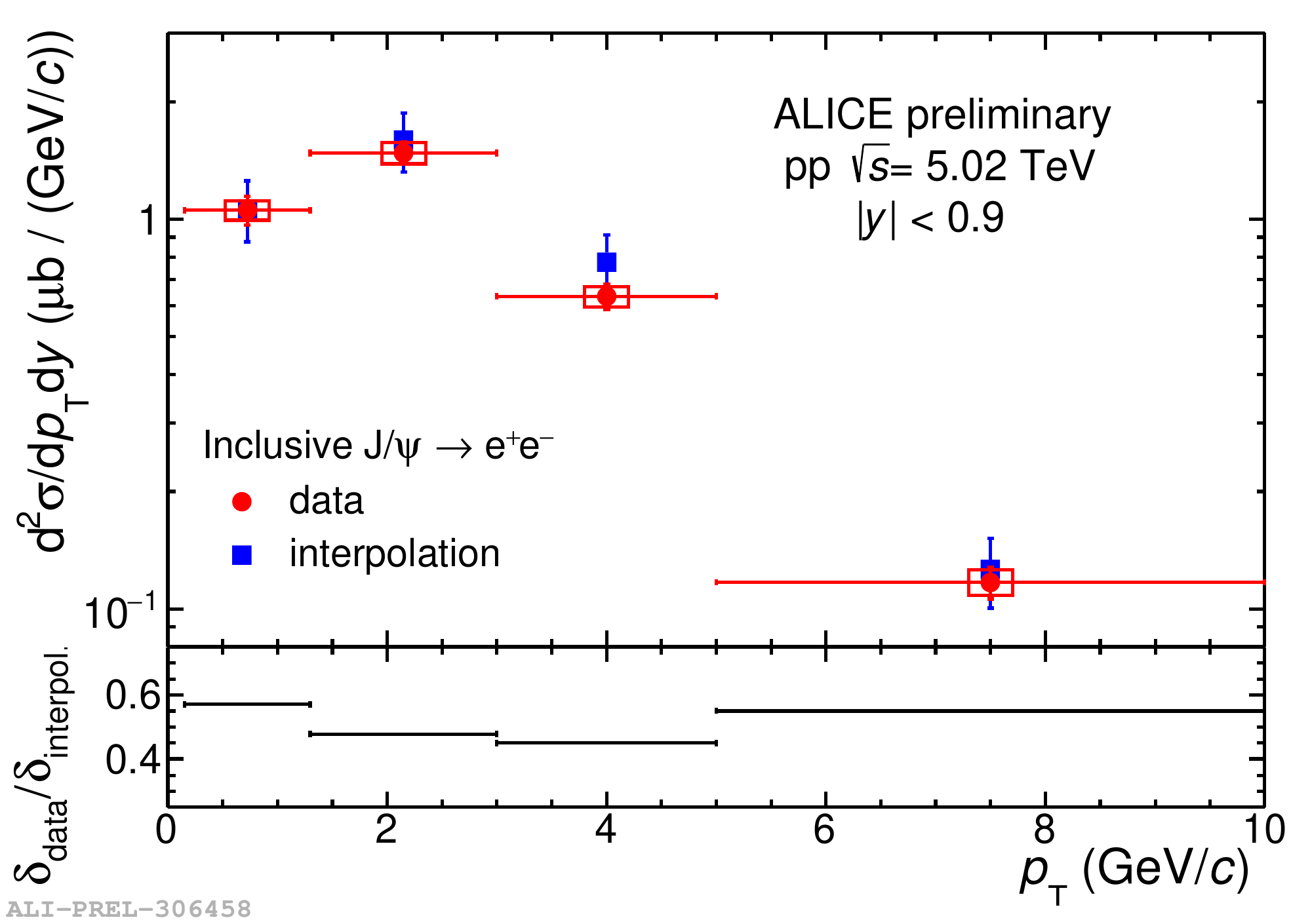}
  \includegraphics[width = 7.3 cm]{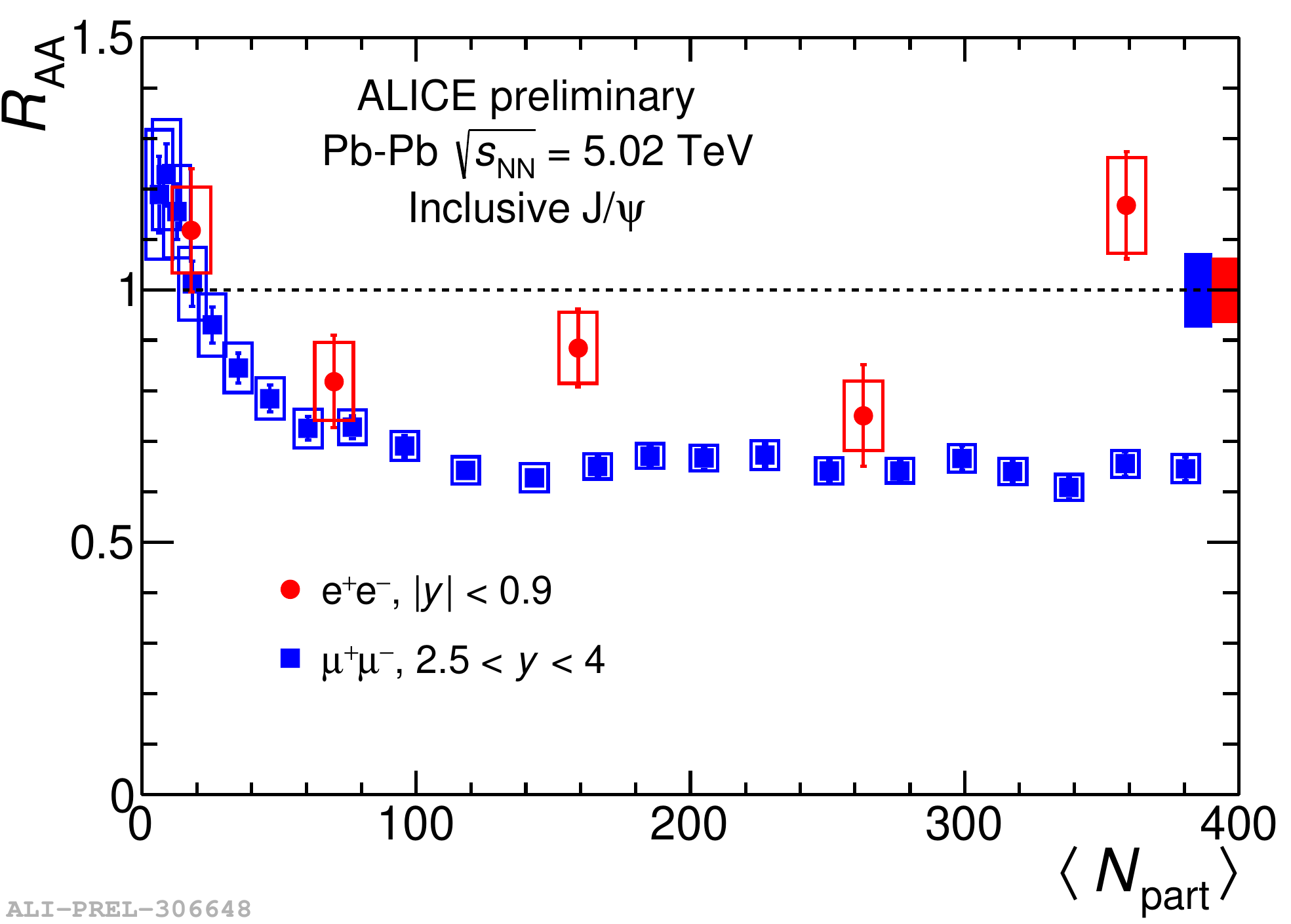}
  \caption{Left: The transverse momentum spectrum measured in pp collisions at $ \s = 5.02$~TeV is compared to that obtained with an interpolation and the ratio of their uncertainties is shown in the lower panel. Right: The nuclear modification factor for $\PbPb$ collisions at $\snn = 5.02$~TeV is shown as a function of centrality for mid- and forward rapidity.}\label{fig:1}
\end{figure}

A precise measurement of the nuclear modification factor relies on the accurate knowledge of the vacuum reference. At mid-rapidity, the analysis of the $\pp$ data taken in 2017 at $\s = 5.02$~TeV replaces the previously used $\jpsi$ reference obtained using an interpolation procedure~\cite{JulianThesis}. In the left panel of Figure~\ref{fig:1}, the measured transverse momentum spectrum is compared to the interpolated spectrum in the upper panel and the ratio of their uncertainties is shown in the lower panel. A reduction of the total uncertainty of a factor of two is achieved by the new measurement while the results agree within errors.

In the right panel of Figure~\ref{fig:1}, the $\jpsi$ nuclear modification factor as a function of centrality for $\PbPb$ collisions at $\snn = 5.02$~TeV is shown. The mid-rapidity result obtained using the measured pp reference is shown together with results measured at forward rapidity~\cite{ALICE_jpsi_PbPb_forw}. While for peripheral collisions the suppression pattern for mid- and forward rapidity seems comparable, for central collisions the nuclear modification factor at mid-rapidity is significantly larger and exceeds unity. In the (re)generation scenario, this can be explained by the larger charm quark density at mid-rapidity which increases the production of $\jpsi$. 

\begin{figure}[t]
  \centering
  \includegraphics[width = 7.1 cm]{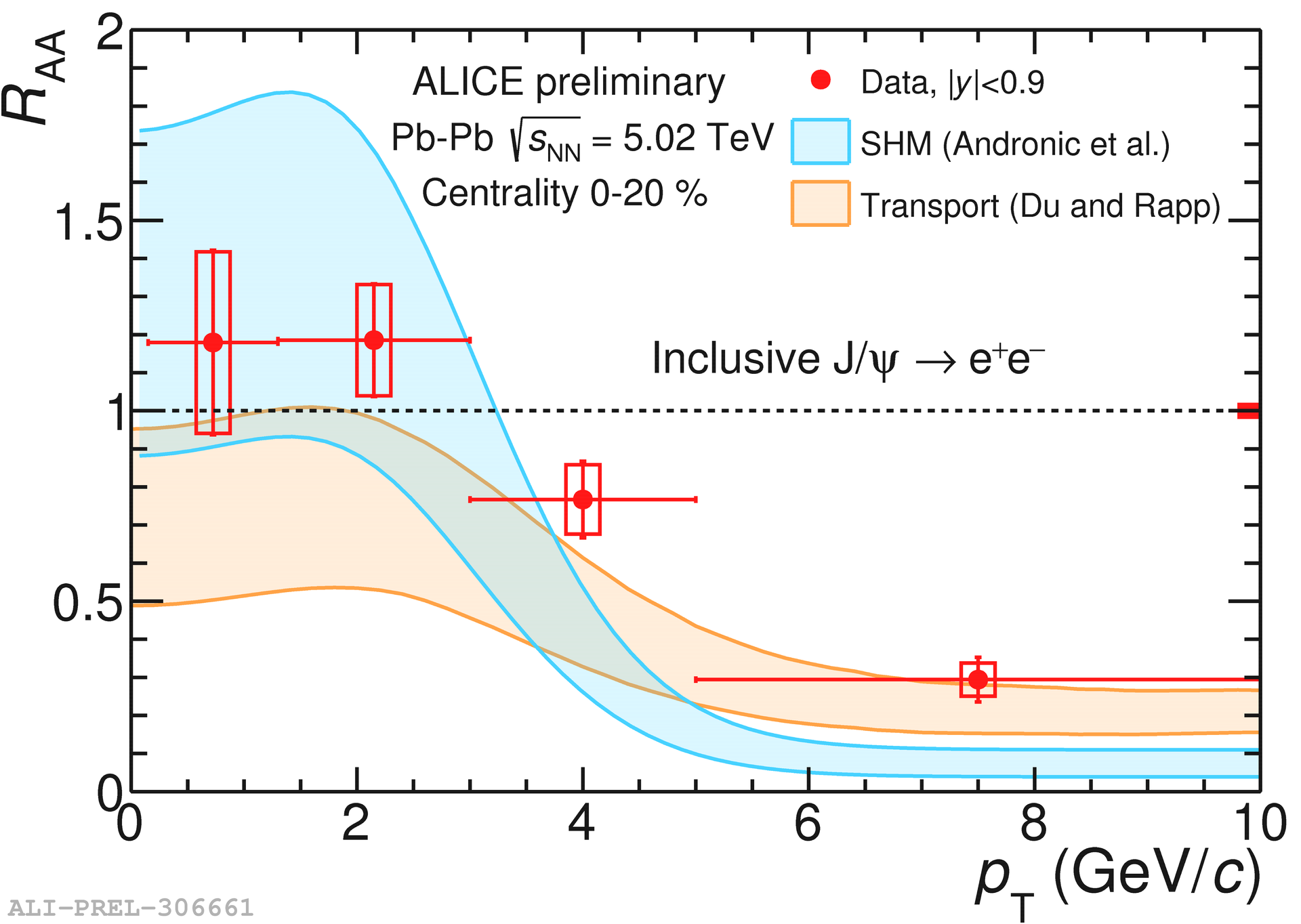}
  \includegraphics[width = 7.9 cm]{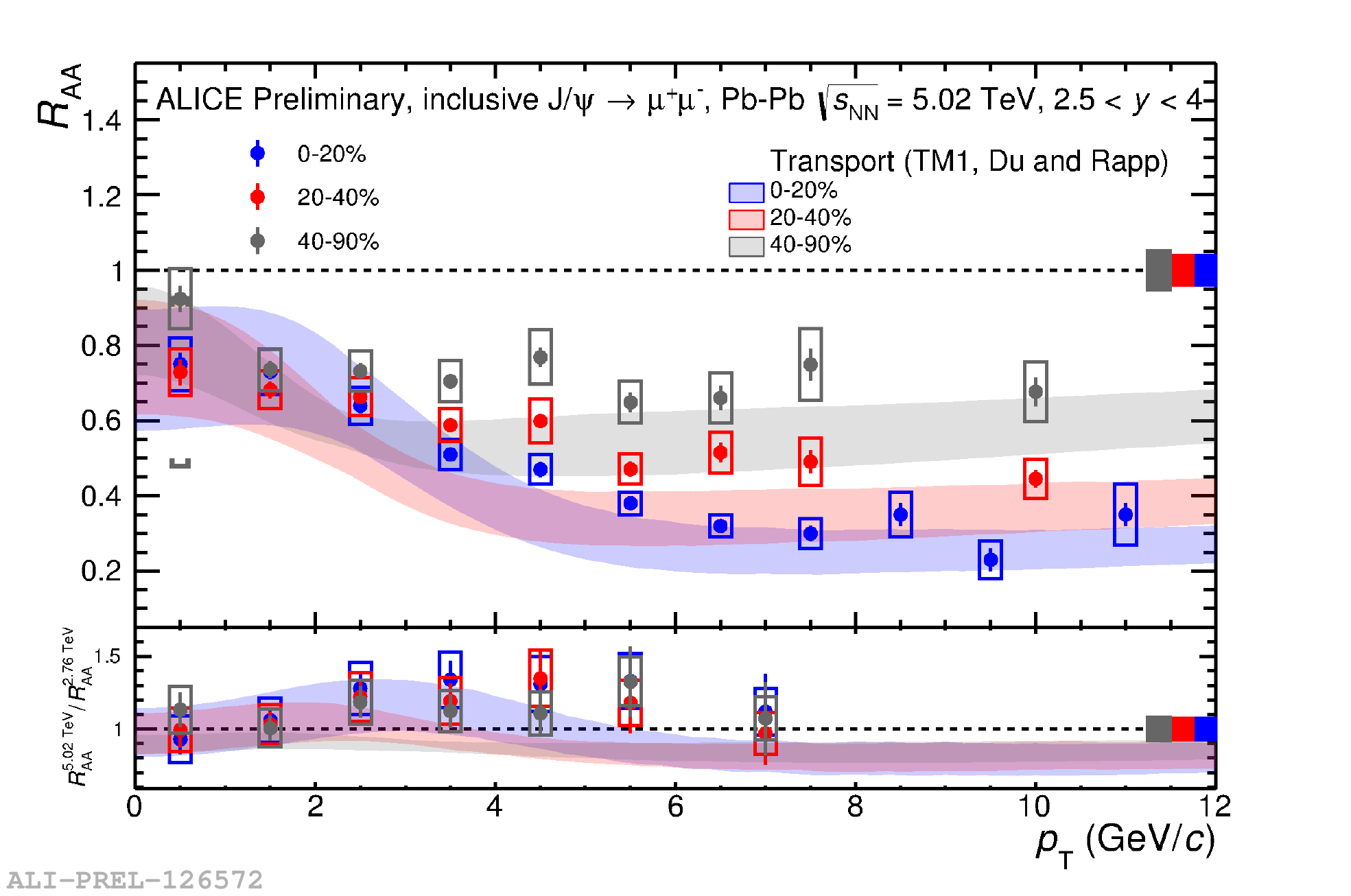}
  \caption{The $\jpsi$ nuclear modification factor as a function of transverse momentum is shown for mid- (left) and forward-rapidity (right) and compared with model calculations~\cite{SHM_charm_2018,Rapp_charm_2015}.}\label{fig:2}
\end{figure}

In Figure~\ref{fig:2}, the $\jpsi$ $\RAA$ as a function of transverse momentum is shown with results from phenomenological models. In the left panel, the $\jpsi$ $\RAA$ at mid-rapidity for the $20$\% most central collisions is compared to the statistical hadronisation model~\cite{SHM_charm_2018} and a transport model~\cite{Rapp_charm_2015}. The statistical hadronisation model describes the nuclear modification factor at low $\pT$ but it falls below data for $\pT \gtrsim 4.5$~\GeVc. The transport model describes the shape of the data although it underestimates the data in the whole $\pt$ range. In the (re)generation picture the increase towards lower $\pT$, as seen in data and models, reflects the phase space where charm quarks are thermalised and consequently form $\jpsi$ by (re)generation which have a lower $\pT$ than primordial $\jpsi$. In the right panel of Figure~\ref{fig:2}, the $\jpsi$ nuclear modification factor at forward rapidity is compared to a transport model~\cite{Rapp_charm_2015} for three different event centralities. The model describes the behaviour of the data quantatively although seems to undershoot the data for $\pt \gtrsim 4.5$~\GeVc.

\begin{figure}[b]
  \centering
  \includegraphics[width = 7.2 cm]{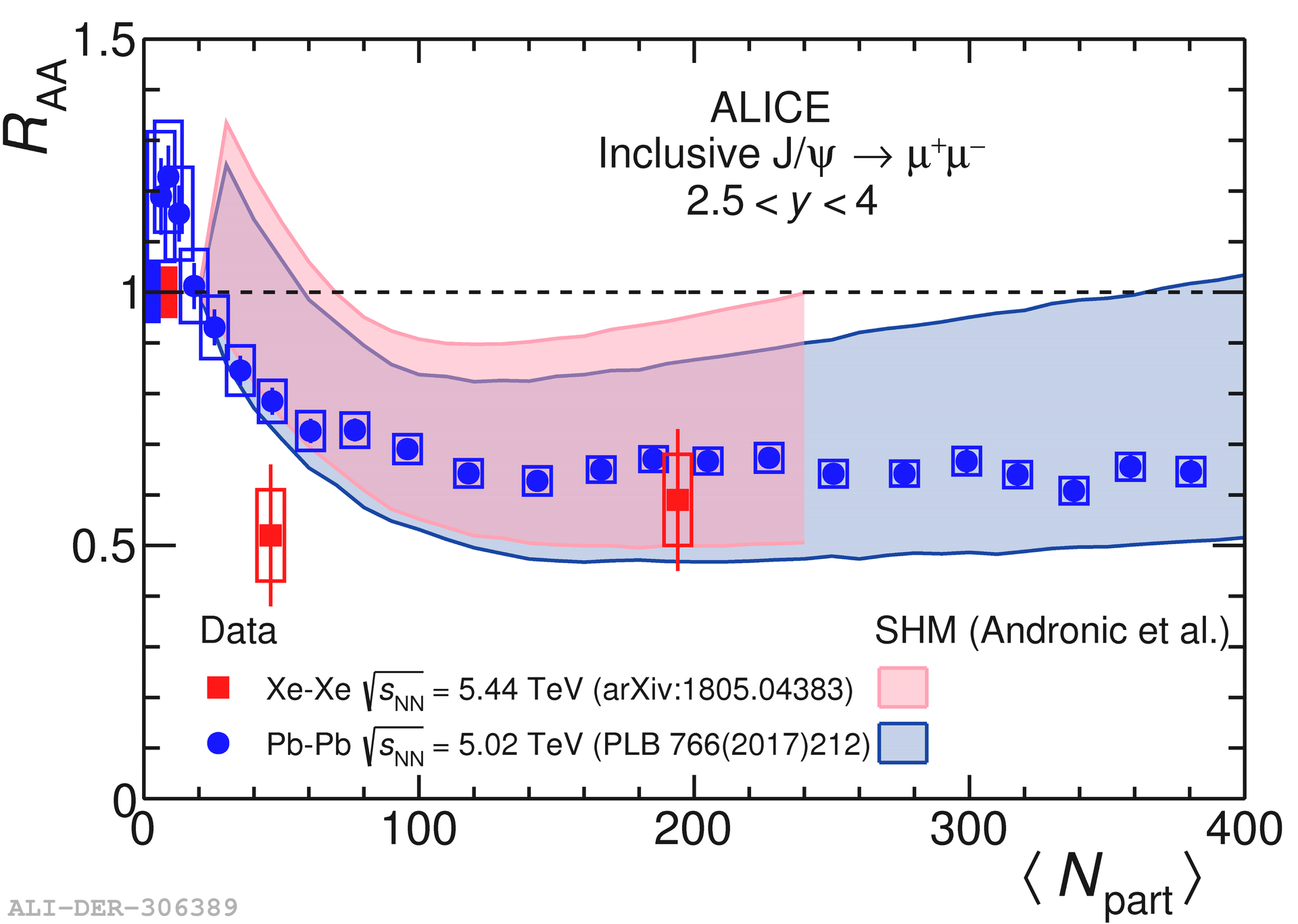}
  \includegraphics[width = 7. cm]{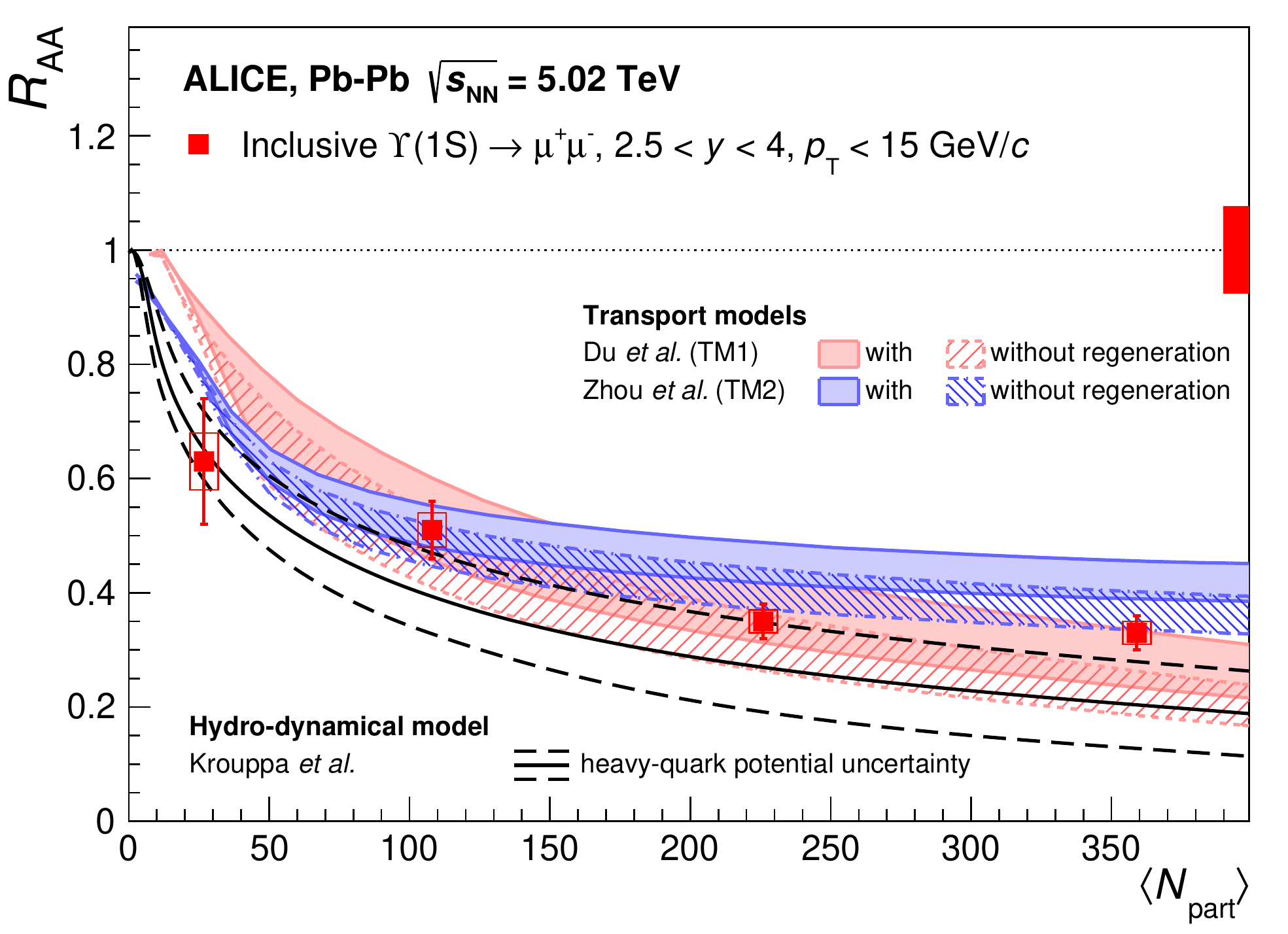}
  \caption{Left: The $\jpsi$ nuclear modification factor as a function of centrality is shown for $\XeXe$ collisions at $\snn = 5.44$~TeV~\cite{ALICE_jpsi_XeXe_forw} and $\PbPb$ collisions~\cite{ALICE_jpsi_PbPb_forw} in comparison with a model~\cite{SHM_charm_2018}. Right: The nuclear modification factor of $\Ups (1S)$ is shown as a function of centrality for $\PbPb$ collisions at $\snn = 5.02$~TeV~\cite{ALICE_ups_PbPb_forw} in comparison with various models~\cite{Du_2017,Zhou_2014,Krouppa_2018}.}\label{fig:3}
\end{figure}

The $\jpsi$ nuclear modification factor for $\XeXe$ collisions at $\snn = 5.44$~TeV as a function of centrality~\cite{ALICE_jpsi_XeXe_forw} is shown in the left panel of Figure~\ref{fig:3} and compared to $\PbPb$ collisions at $\snn = 5.02$~TeV~\cite{ALICE_jpsi_PbPb_forw}. The nuclear modification factor of the two collision systems is of the same magnitude for a given $\Npart$. This behaviour is also reflected both by the transport and statistical hadronisation models. 

In the right panel of Figure~\ref{fig:3}, the $\Ups(1\text{S})$ nuclear modification factor as a function of centrality is shown for $\PbPb$ collisions at $\snn = 5.02$~TeV~\cite{ALICE_ups_PbPb_forw}. A strong suppression towards central collisions is observed, reminiscent of the $\jpsi$ suppression at SPS or RHIC energies~\cite{Arnaldi_2016}. Data is compared with two transport models~\cite{Du_2017,Zhou_2014} and a hydro-dynamical model~\cite{Krouppa_2018}. The transport models are shown including or excluding (re)generation as production mechanism. It can be seen that the (re)generation component is moderate and the data does not allow for the discrimination between the two scenarios.

%
%
Summarising, ALICE measured the production and nuclear modification of quarkonium at mid- and forward rapidity in \PbPb and \XeXe collisions. The results suggest that (re)generation is the dominant contribution to the production of $\jpsi$, but contributes only marginally to the production of $\Ups(1\text{S})$ in heavy-ion collisions at LHC collision energies. 

The current precision of the models allows only a weak discrimination of the different phenomenological approaches. A more precise determination of the initial heavy-quark production cross section and the recent $\PbPb$ data taken in 2018 will put stronger constraints on the models.


\begin{thebibliography}{99}
\bibitem{Brambilla2011}
  N.~Brambilla et al., Eur. Phys. C71 (2011) 1534, \href{https://arxiv.org/abs/1010.5827}{arXiv:1010.5827 [hep-ph]}
\bibitem{Andronic2016}
  A.~Andronic et al., Eur. Phys. C76 (2016) 107, \href{https://arxiv.org/abs/1506.03981}{arXiv:1506.03981 [nucl-ex]}
\bibitem{Bodwin1995}
  G.~T.~Bodwin, E.~Braaten and G.~P.~Lepage, Phys. Rev. D51 (1995) 1125 [Erratum-ibid. D55 (1997) 5853], \href{https://arxiv.org/abs/hep-ph/9407339}{arXiv:hep-ph/9407339}
\bibitem{Andronic2007}
  A.~Andronic, P.~Braun-Munzinger, K.~Redlich and J.~Stachel, Nucl. Phys. A789 (2007) 334, \href{https://arxiv.org/abs/nucl-th/0611023}{arXiv:nucl-th/0611023}
\bibitem{MatsuiSatz:1986}
  T.~Matsui and H.~Satz, Phys. Lett. B178 (1986) 416
\bibitem{Arnaldi_2016}
  R.~Arnaldi, Nucl. Phys. A956 (2016) 128, \href{https://arxiv.org/abs/1604.03139}{arXiv:1604.03139 [hep-ex]} 
\bibitem{BraunMunzinger:2000px}
  P.~Braun-Munzinger and J.~Stachel, Phys. Lett. B490 (2000) 196, \href{https://arxiv.org/abs/nucl-th/0007059}{arXiv:nucl-th/0007059}
\bibitem{Thews:2000rj}
  R.~L.~Thews, M.~Schroedter and J.~Rafelski, Phys. Rev. C63 (2001) 054905, \href{https://arxiv.org/abs/hep-ph/0007323}{arXiv:hep-ph/0007323}
\bibitem{ZhaoRapp:2011}
  X.~Zhao and R.~Rapp, Nucl. Phys. A859 (2011) 114, \href{https://arxiv.org/abs/1102.2194}{arXiv:1102.2194 [hep-ph]}
\bibitem{Ferreiro:2012rq}
  E.~G.~Ferreiro, Phys. Lett. B731 (2014) 57, \href{https://arxiv.org/abs/1210.3209}{arXiv:1210.3209 [hep-ph]}
\bibitem{ALICE_XeXe_pt}
  ALICE Collaboration, submitted to Phys. Lett. B, \href{https://arxiv.org/abs/1805.04399}{arXiv:1805.04399 [nucl-ex]}
\bibitem{ALICE_XeXe_rapidity}
  ALICE Collaboration, submitted to Phys. Lett. B, \href{https://arxiv.org/abs/1805.04432}{arXiv:1805.04432 [nucl-ex]}
\bibitem{ALICE_XeXe_flow}
  ALICE Collaboration, Phys.Lett. B784 (2018) 82, \href{https://arxiv.org/abs/1805.01832}{arXiv:1805.01832 [nucl-ex]}
\bibitem{ALICE_detector2008}
  ALICE Collaboration, JINST 3 (2008) S08002
\bibitem{ALICE_detector2014}
  ALICE Collaboration, Int. J. Mod. Phys. A 29 (2014) 1430044, \href{https://arxiv.org/abs/1402.4476}{arXiv:1402.4476 [nucl-ex]}
\bibitem{JulianThesis}
  J. Book, \href{https://www.uni-frankfurt.de/54952831/PHD_THESIS_JULIAN_BOOK.pdf}{Doctoral thesis}, University of Frankfurt, 2015
\bibitem{ALICE_jpsi_PbPb_forw}
  ALICE Collaboration, Phys. Lett. B766 (2017) 212, \href{https://arxiv.org/abs/1606.08197}{arXiv:1606.08197 [nucl-ex]}
\bibitem{SHM_charm_2018}
  A.~Andronic, P.~Braun-Munzinger, M.~K.~K\"{o}hler and J.~Stachel, \href{https://arxiv.org/abs/1807.01236}{arXiv:1807.01236 [nucl-th]}
\bibitem{Rapp_charm_2015}
  X.~Du and R.~Rapp, Nucl. Phys. A943 (2015) 147, \href{https://arxiv.org/abs/1504.00670}{arXiv:1504.00670 [hep-ph]}
\bibitem{ALICE_jpsi_XeXe_forw}
  ALICE Collaboration, Phys. Lett. B785 (2018) 419, \href{https://arxiv.org/abs/1805.04383}{arXiv:1805.04383 [nucl-ex]}
\bibitem{ALICE_ups_PbPb_forw}
  ALICE Collaboration, Submitted to Phys. Lett. B, \href{https://arxiv.org/abs/1805.04387}{arXiv:1805.04387 [nucl-ex]}
\bibitem{Du_2017}
  X.~Du, M.~He and R.~Rapp, Phys. Rev. C96 (2017) 054901, \href{https://arxiv.org/abs/1706.08670}{arXiv:1706.08670 [hep-ph]}
\bibitem{Zhou_2014}
  K.~Zhou, N.~Xu and P.~Zhuang, Nucl. Phys. A931 (2014) 654, \href{https://arxiv.org/abs/1408.3900}{arXiv:1408.3900 [hep-ph]}
\bibitem{Krouppa_2018}
  B.~Krouppa, A.~Rothkopf and M.~Strickland, Phys. Rev. D97 (2018) 016017, \href{https://arxiv.org/abs/1710.02319}{arXiv:1710.02319 [hep-ph]}
\end{thebibliography}
\end{document}